\documentclass[a4paper,11pt]{article}
\usepackage{pos}
\usepackage{graphicx,epsfig,dcolumn,bm,epic,eepic,float}
\usepackage{amsmath}
\usepackage[T1]{fontenc}
\usepackage[utf8]{inputenc}
\usepackage{mathtools} 
\usepackage{pifont}
\usepackage{amsmath}
\usepackage{latexsym}
\usepackage{color}
\usepackage{cancel}
\usepackage{scrextend}
\usepackage{amsmath}
\usepackage{autobreak}
\usepackage{amsthm}
\usepackage{eucal}
\usepackage{mathrsfs}
\usepackage[bottom]{footmisc}
\usepackage{makeidx,shortvrb,latexsym}
\usepackage{epstopdf}
\usepackage{mathtools}
\usepackage{ragged2e}
\usepackage[T1]{fontenc}
\newcommand {\be}{\begin{equation}}
\newcommand {\ee}{\end{equation}}
\newcommand {\ba}{\begin{eqnarray}}
\newcommand {\ea}{\end{eqnarray}}
\newcommand {\bea}{\begin{eqnarray}}
\newcommand {\eea}{\end{eqnarray}}

\pdfoutput=1

\DeclareMathAlphabet{\mathcal}{OMS}{cmsy}{m}{n}

\title{Higgs-Sector Predictions from Maximally Symmetric multi-Higgs Doublet Models}

\author[a,b]{Neda Darvishi}
\author[c]{M.R. Masouminia}
\author*[d,e]{Apostolos Pilaftsis}

\affiliation[a]{Department of Physics and Astronomy, Michigan State University,\\
East Lansing, MI 48824, USA}

\affiliation[b]{Institute of Theoretical Physics, Chinese Academy of Sciences,\\
Beijing 100190, China}

\affiliation[c]{Institute for Particle Physics Phenomenology, Durham University,\\
Durham DH1 3LE, United Kingdom}

\affiliation[d]{Department of Physics and Astronomy,
University of Manchester,\\
Manchester M13 9PL, United Kingdom}

\affiliation[e]{Theoretical Physics Department, CERN, CH-1211 Geneva 23, Switzerland}

\emailAdd{neda.darvishi@itp.ac.cn}
\emailAdd{mohammad.r.masouminia@durham.ac.uk}
\emailAdd{apostolos.pilaftsis@manchester.ac.uk}

\abstract{Maximally Symmetric $n$-Higgs Doublet Models (MS-$n$HDMs)
  define very economic settings that enable sharp Higgs-sector
  predictions beyond the Standard Model (SM) potentially testable at
  high-energy colliders. The scalar potential of a MS-$n$HDM obeys an
  $\mathrm{Sp(2}n)$ symmetry, which is softly broken by bilinear
  scalar masses and explicitly by hypercharge and Yukawa couplings
  through renormalisation-group effects. The $\mathrm{Sp(2}n)$ also
  ensures natural SM alignment and allows\- for quartic coupling
  unification up to the Planck~scale. As typical examples, we consider
  maximally symmetric realisations of the Type-II 2HDM and the Type-V
  3HDM. We show how in terms of a few input parameters, definite
  predictions for the entire scalar mass spectrum of the MS-2HDM and
  MS-3HDM are obtained, including the SM-like Higgs-boson couplings to
  the gauge bosons and fermions.}

\FullConference{7th Symposium on Prospects in the Physics of Discrete
  Symmetries (DISCRETE 2020-2021)\\ 
  29th November - 3rd December 2021, Bergen, Norway}

\begin{document}
\maketitle 

\section{Introduction}
\label{sec:intro} 

An interesting class of Higgs-sector extensions is the one that augments the SM with~$n$~Higgs doublets
$\Phi_{i}^\mathsf{T}=\, \left(
 \begin{matrix} \Phi_i^+, \, \Phi_i^0
 \end{matrix}\right)^{\mathsf{T}}$ (with $i=2, \cdots, n$), usually called the
$n$-Higgs Doublet Model ($n$HDM). The general $n$HDM potential may be conveniently expressed as follows~\cite{Botella:1994cs}:
\begin{equation}
  V=\, -\displaystyle\sum_{i,j=1}^{3}\, m^2_{ij}\, ( \Phi_i^{\dagger} \Phi_j) \,+\, \ {1\over 2} \displaystyle\sum_{i,j\, k,l=1}^{3}\, 
  \lambda_{ijkl}\, ( \Phi^{\dagger}_i \Phi_j)( \Phi^{\dagger}_k \Phi_l),
  \label{pot-nhdm}
\end{equation}
with $\lambda_{ijkl}\,=\, \lambda_{klij}$. In the $n$HDM, the couplings of the
SM-like Higgs boson to the EW gauge bosons ($Z,\,W^\pm$) must
resemble those predicted by the SM, so as to be in agreement with the
current Higgs signals at the LHC. This is only possible within the
so-called SM alignment
limit~\cite{Ginzburg:1999fb,Chankowski:2000an,Delgado:2013zfa,Carena:2013ooa,BhupalDev:2014bir,Bernon:2015qea,Darvishi:2017bhf,Benakli:2018vqz,Lane:2018ycs,Darvishi:2020paz}.

The potential of $n$HDMs contains a large number of
${\mathrm{SU(2)}}_L$-preserving accidental symmetries as subgroups of
the symplectic group ${\mathrm{Sp}}(2n)$. This maximal symmetry group
plays an instrumental role in classifying accidental symmetries that
may occur in the scalar potentials of $n$HDMs and $n$HDM-Effective
Field Theories with higher-order
operators~\cite{Pilaftsis:2016erj,Darvishi:2019dbh,Darvishi:2020teg,Birch-Sykes:2020btk}. Thus
far, this classification has been done for: (i) the
2HDM~\cite{Battye:2011jj,Pilaftsis:2011ed}, (ii) the 2HDM Effective Field Theory for
higher-order operators up to dimension-6 and
dimension-8~\cite{Birch-Sykes:2020btk}, and (iii)~the
3HDM~\cite{Ivanov:2012ry,Darvishi:2019dbh,Darvishi:2021txa}. From
these symmetries, only a few possess the desirable property of natural
SM alignment~\cite{Pilaftsis:2016erj,Darvishi:2020teg} \bea
{\rm (i)}\,\, \mathrm{Sp}(2n), \qquad{\rm (ii)}\, \,
\mathrm{SU}(n)_{\rm HF}, \qquad {\rm (iii)}\,\, \mathrm{SO}(n)_{\rm
  HF} \times {\rm CP}.
\label{NAS}
\eea Evidently, the $\mathrm{Sp}(2n)$-invariant $n$HDM is the most
economic setting that realises naturally such an alignment from the
general class of $n$HDMs.  The other two class of models, based on
$\mathrm{SU}(n)_{\mathrm{HF}}$ and
$\mathrm{SO}(n)_{\rm HF} \times {\rm CP}$ groups, require one and two
extra parameters, respectively, as compared to the Maximally Symmetric
$n$HDM (MS-$n$HDM) that implements the $\mathrm{Sp}(2n)$ group. These
additional theoretical parameters make these symmetric models less
predictive, while spoiling the interesting feature of quartic coupling
unification.  In the following sections, after discussing the basic
feature of Type II-2HDM and Type V-3HDM, we will focus on MS-$n$HDM
with $n=2,3$ Higgs doublets.

\section{The Higgs Spectrum of the Type-II 2HDM}

The Higgs sector of the 2HDM is described by two scalar doublets $\Phi_{1}$ and $\Phi_{2}$.
Performing the usual linear expansion of these scalar doublets about their VEVs, we may express them as
\begin{equation}
\Phi_i \  =\ \left(
 \begin{matrix}
 \phi_i^+ \\
 {1 \over \sqrt{2}} (v_i + \phi_i + i\chi_i)
 \end{matrix}
 \right)\;,
 	\label{vev-ijk}
\end{equation}
with $i=1,2$. In the case of CP-conserving Type-II 2HDM, both
scalar doublets receive real and nonzero vacuum
expectation values (VEVs). In detail, we have
$\langle \Phi_1^0\rangle= v_1/\sqrt{2}$ and
$\langle \Phi_2^0 \rangle = v_2/\sqrt{2}$, where $\tan\beta =v_2/v_1$ and $v \equiv (v_1^2+v_2^2)^{1/2}$ forms the VEV of the SM Higgs doublet. 

Our first step is to transform all scalar fields from a weak basis with generic choice of vacua to the so-called Higgs basis~\cite{Georgi:1978ri,Haber:2006ue}, where only one Higgs doublet acquires the SM VEV. This can be achieved by virtue of a common orthogonal transformation that involves
the mixing angles $\beta$, i.e.
\begin{equation}
\begin{pmatrix} H_{1}\\ H_{2}  \end{pmatrix}
=\mathcal{R}(\beta) \begin{pmatrix} \phi_1 \\ \phi_2 \end{pmatrix},
\quad 
\begin{pmatrix} G^0 \\ a \end{pmatrix}
=\mathcal{R}(\beta) \begin{pmatrix} \chi_1 \\ \chi_2 \end{pmatrix},
\quad
\begin{pmatrix} G^{\pm} \\ h^{\pm} \end{pmatrix}
= \mathcal{R}(\beta) \begin{pmatrix} \phi_1^{\pm} \\ \phi_2^{\pm} \end{pmatrix},
\end{equation}
where the two-dimensional rotational matrix is
\begin{align}
R(\beta)&= \left(
\begin{array}{ccccc}
\cos \beta& \sin \beta\\ 
-\sin \beta&\cos \beta
     \end{array}
     \right).
\end{align}

After spontaneous symmetry breaking (SSB), the $Z$ and $W^{\pm}$ gauge bosons become massive after absorbing in the unitary gauge the three would-be Goldstone bosons $G^0$ and $G^{\pm}$, respectively~\cite{Goldstone1}. Consequently, the 2HDM can account for only five physical scalar states: two
CP-even scalars ($h$,$H$), one CP-odd scalar $(a)$ and two charged
bosons ($h^{\pm}$). The masses of the $a$ and $h^\pm$  scalars with $s_{\beta} \equiv \sin{\beta}$ and $c_{\beta} \equiv \cos{\beta}$ are given by
\bea
M_a^{2} \, =\, M_{h^{\pm}}^{2} + {v^2 \over 2} (\lambda_4 \,-\, \lambda_5),\quad
M_{h^{\pm}}^{2} \, =\, {m_{12}^{2} \over c_{\beta} s_{\beta}}\,-\, {v^{2} \over 2} (\lambda_4 \,+\, \lambda_5)
 + {v^{2} \over 2 c_{\beta} s_{\beta} } (\lambda_6 c_{\beta}^{2} 
 + \lambda_7 s_{\beta}^{2}).
\eea
The masses of the two~CP-even scalars, $h$ and $H$ may be obtained by
diagonalising the CP-even mass matrix~$\mathcal{M}^{2}_S$,
\begin{align}
\overline{\mathcal{M}}_{\rm S}^2 =
\left(
\begin{array}{cc}
{M}_{H}^2 & 0 
\\
0 & {M}_{h} ^2
\end{array}
\right)=\ R(\alpha)\,
{\mathcal{M}}_{\rm S}^2 \,{R(\alpha)}^{\sf T}=\ R(\alpha)\,
\left(
\begin{matrix}
{A}\,&\, {C} \\
{C}\,&\, {B}
 \end{matrix}\right) \,{R(\alpha)}^{\sf T}.
\label{mass-wtom-2hdm}
\end{align}
The details of the above matrix are presented in \cite{Darvishi:2019ltl}.
Additionally, the mixing angle $\alpha$ may be determined by $\tan 2\alpha\,=\,{2 C/ A-B}$.

To this extend, one may obtain the SM-normalised couplings of the CP-even scalars, $h$ and $H$, to the gauge bosons ($V \,=\, W^{\pm}, Z$) as follows:
\begin{equation}
g_{hVV} = \sin (\beta - \alpha), \qquad g_{HVV} = \cos (\beta - \alpha).
\end{equation}
Therefore, there are two scenarios to realise
the SM alignment limit: (i) SM-like ~$H$ scenario with $\cos(\beta-\alpha) \to 1$, and (ii) SM-like ~$h$ scenario $\sin(\beta-\alpha) \to 1$.
Here, we consider the first scenario with $\cos(\beta-\alpha) \to 1$.

In the Higgs basis, the CP-even mass matrix takes on the form
 \begin{eqnarray}
{\widehat{\mathcal{M}}_S}^{\,2} & \,=\, &  \left(
 \begin{matrix}
 \widehat{A} & \widehat{C} \\
 \widehat{C} & \widehat{B}
 \end{matrix}
 \right) \,=\, R(\beta) {\mathcal{M}_S}^{2}R(\beta)^{\sf T}\,,
 \label{MS2}
\end{eqnarray}
with
\begin{eqnarray}
\label{Chat}
\widehat{A} & = & 2 v^{2} \left[ c_{\beta}^4 \lambda_1 + s_{\beta}^{2} c_{\beta}^{2} \lambda_{345}
 + s_{\beta}^4 \lambda_2 + 2 c_{\beta} s_{\beta} \left( c_{\beta}^{2} \lambda_6 
 + s_{\beta}^{2} \lambda_7 \right) \right], 
\nonumber \\
\widehat{B} & = & M_a^{2} + \lambda_5 v^{2} + 2 v^{2} \left[ s_{\beta}^{2} c_{\beta}^{2} \left(
 \lambda_1 + \lambda_2 - \lambda_{345} \right) - c_{\beta} s_{\beta}  \left(
 c_{\beta}^{2} - s_{\beta}^{2} \right) \left(\lambda_6-\lambda_7 \right) \right],
 \\ 
\widehat{C} & = & v^{2} \left[ s_{\beta}^3 c_{\beta} \left( 2 \lambda_2 - \lambda_{345} \right)
 - c_{\beta}^3 s_{\beta} \left( 2 \lambda_1 - \lambda_{345} \right)
 + c_{\beta}^{2} \left( 1 - 4 s_{\beta}^{2} \right) \lambda_6
 + s_{\beta}^{2} \left(4 c_{\beta}^{2} - 1 \right) \lambda_7 \right].
\nonumber 
\end{eqnarray}
Thereby, in the SM alignment limit $\beta=\alpha$
the mass parameter $\widehat{A}$ becomes equal to~$M_H^2$, while the parameters $\widehat{C}$ vanishes. Thus, in the limit $\widehat{C} \to 0$, we obtain the following conditions:
 \bea
\lambda_{1}=\lambda_{2}={(\lambda_{3}+\lambda_{4}+\lambda_{5})/2}, \quad  \lambda_{6}=\lambda_{7}=0,
\label{2hdm-Cto0}
\eea
where by imposing symmetries identified in \eqref{NAS} with $n=2$ the above relationships between the quartic couplings may be fulfilled naturally.

Moreover, the deviation of the $H$-boson couplings from their SM values in terms of the light-to-heavy~scalar-mixing parameter $\theta_\mathcal{S}\equiv\widehat{C}/\widehat{B}$ are given by
\bea
g_{HVV} \simeq 1-\theta_\mathcal{S}^{2}/2,
\qquad g_{hVV} \simeq -{\theta_\mathcal{S}}.
\eea
where the mixing parameter $\theta_\mathcal{S}$
vanishes in the exact SM alignment limit as
$\alpha \to \beta$.  Accordingly, we may derive approximate analytic expressions for the 
$h$- and $H$-boson couplings to up- and down-type quarks to leading order in ~$\theta_\mathcal{S}$, as
\begin{align}
g_{Huu}&\simeq  1+{t^{-1}_\beta}\,{\theta_\mathcal{S}}, \qquad \qquad
 g_{Hdd} \simeq 1-{\theta_\mathcal{S}}\,{t_\beta}, \\ \nonumber
 g_{huu}&\simeq -{\theta_\mathcal{S}}+{t^{-1}_\beta}, \qquad \qquad
 g_{hdd} \simeq -{\theta_\mathcal{S}}-{t_\beta} .
   \label{mfc}
\end{align}
In the SM alignment limit, we have $g_{Huu}\,\to\,1$ and
$g_{Hdd} \to$ 1.  Obviously, any deviation of these couplings from their SM values is governed by quantities $\tan\beta$ and $\theta_\mathcal{S}$.

\section{The Higgs Spectrum of the Type-V 3HDM}

In the case of CP-conserving Type-V 3HDM, the VEVs of scalar doublets,
$\Phi_i$ $(i=1,2,3)$ may be given by   
\bea
v_1 \,=\, v \cos\beta_1 \cos\beta_2, \, \quad 
v_2 \,=\, v \sin\beta_1 \cos\beta_2, \,\quad 
v_3 \,=\, v \sin\beta_2.
\eea
Equivalently, the VEVs may be defined by the two ratios, $\tan\beta_1 = v_2/v_1$ and $ \tan\beta_2 = v_3/\sqrt{v_1^2+ v_2^2}$,  given the constraint that $v \equiv \sqrt{v_1^2 + v_2^2 + v_3^2}$ is the SM VEV.

For later convenience, we proceed as in~\cite{Das:2019yad} and define the three-dimensional rotational matrices about the individual axes $z,\, y$ and $x$ as follows:
\begin{align}
 R_{12}(\alpha)= \left( \begin{array}{ccccc}
  \cos\alpha & \sin\alpha&0\\ 
  -\sin\alpha&\cos\alpha& 0\\
  0&0& 1
     \end{array}
     \right),\,\,
R_{13}(\beta) = \left(
\begin{array}{ccccc} 
\cos\beta& 0& \sin\beta\\
0&1 & 0 \\
-\sin\beta & 0&\cos\beta\end{array}
     \right),\,\,
R_{23}(\gamma) = \left(
\begin{array}{ccccc}  
 1 & 0 & 0\\
 0& \cos\gamma&  \sin\gamma\\
 0& -\sin\gamma& \cos\gamma
  \end{array}
\right).
\end{align}

In the case of 3HDM, the two mixing angles $\beta_1$ and $\beta_2$ are required to transform all scalar fields to the Higgs basis with $\mathcal{O}_\beta\equiv R_{13}(\beta_2)R_{12}(\beta_1)$, i.e.
\bea
\begin{pmatrix}
H_{1}\\ H_2 \\ H_3
\end{pmatrix}
=
\mathcal{O}_\beta
\begin{pmatrix}
\phi_1 \\ \phi_2 \\ \phi_3
\end{pmatrix},
\qquad 
\begin{pmatrix}
G^0 \\ \eta_1 \\ \eta_2
\end{pmatrix}
=
\mathcal{O}_\beta
\begin{pmatrix}
\chi_1 \\ \chi_2 \\ \chi_3
\end{pmatrix},
\qquad
\begin{pmatrix}
G^{\pm} \\ \eta_1^{\pm} \\ \eta_2^{\pm}
\end{pmatrix}
=
\mathcal{O}_\beta
\begin{pmatrix}
\phi_1^{\pm} \\ \phi_2^{\pm} \\ \phi_3^{\pm}
\end{pmatrix}.
\eea
Therefore, after SSB the model exhibits nine scalar mass eigenstates: (i)~three CP-even scalars ($H,h_{1},h_{2}$), (ii)~two CP-odd scalars $(a_{1},a_{2})$, and (iii)~four charged scalars~($h_{1}^{\pm},h_{2}^{\pm}$).
 
In the Higgs basis, spanned by $\{\eta_{1,2}\}$ and $\{\eta^\pm_{1,2}\}$, the CP-odd and charged scalar mass matrices reduce to the $2\times 2$ matrices given by 
\begin{align}
  \mathcal{M}^2_{\text{P}}=
  \left(
  \begin{array}{ccc}
  M^2_{\text{P},22} &  M^2_{\text{P},23}
\\
  M^2_{\text{P},32} &  M^2_{\text{P},33}
\\
\end{array}
\right), \qquad
\mathcal{M}^2_{\pm}=
\left(
\begin{array}{ccc}
  M^2_{{\pm},22} &  M^2_{{\pm},23}
\\
  M^2_{{\pm},32} &  M^2_{{\pm},33}
  \\
\end{array}
  \right),
    \label{eq:MatrixPC}
\end{align}
where the details of all elements are given in \cite{Darvishi:2021txa}. 
Upon diagonalisation of the mass matrices given in~\eqref{eq:MatrixPC}, the masses of the two CP-odd scalars~$(a_{1},a_{2})$ and the four charged Higgs bosons ~($h_{1}^{\pm},h_{2}^{\pm}$) may be computed as 
\begin{align}
  M_{a_1,a_2}^2 &=  {1 \over 2}\bigg[\,M_{{\rm P}, 22}^2 + M_{{\rm P},33}^2 \, \mp \, \big[(M_{{\rm P}, 22}^2- M_{{\rm P},33}^2)^2+4 M_{{\rm P},23}^4\big]^{1/2}\,\bigg]\,,\\
  M^2_{h_1^{\pm},h_2^{\pm}} &=  {1\over 2}\bigg[\,M_{\pm, 22}^2 + M_{\pm,33}^2 \, \mp \, \big[{(M_{\pm,22}^2- M_{\pm,33}^2)^2+4 M_{\pm,23}^4}\big]^{1/2}\,\bigg]\,.
\end{align}
In addition, the mixing angles $\rho$ and $\sigma$ may be given by
\bea
  \tan{2\rho}\,=\,{2 \,M_{{\rm P},23} ^2\over M_{{\rm P},22} ^2-M_{{\rm P},33}^2}, \quad \tan{2\sigma}\,=\,{2\, M_{\pm,23} ^2 \over M_{\pm,22} ^2-M_{\pm,33}^2}.
\eea

Finally, the masses for the three CP-even scalars, $H$, $h_1$ and $h_2$ with mass ordering $M_H \le M_{h_1} \le M_{h_2}$ can be evaluated by diagonalising the squared mass matrix ${\mathcal{M}}^2_{\rm S}$, expressed in the general weak basis $\{\phi_{1,2,3}\}$, by employing the orthogonal matrix $\mathcal{O}=\,R_{23}(\alpha)\, R_{13}(\alpha_2)\, R_{12}(\alpha_1)$, as
\begin{align}
\overline{\mathcal{M}}_{\rm S}^2 \,=\,
\left(
\begin{array}{ccc}
{M}_{H}^2 & 0 & 0
\\
0 & {M}_{h_1} ^2& 0
\\
0 & 0 &{M}_{h_2}^2
\\
\end{array}
\right)\,=\,\mathcal{O}\,
{\mathcal{M}}_{\rm S}^2 \,\mathcal{O}^{\sf T}\,=\,\mathcal{O}\,
\left(
\begin{array}{ccc}
A & {C}_{1} & {C}_{2}
\\
{C}_{1} & B_1& {C}_{3}
\\
{C}_{2} & {C}_{3} &B_2
\\
\end{array}
\right) \,\mathcal{O}^{\sf T}.
\label{mass-wtom}
\end{align}
where the analytic form of all its entries is given in \cite{Darvishi:2021txa}. 

In the CP-conserving 3HDM, the SM-normalised couplings of the SM-like Higgs boson to
the EW gauge bosons ($V = Z,W^{\pm}$) are calculated to be:
\begin{align}
    \label{eq:gHVV-3hdm}
  g_{HVV}\ =&\ \cos{\alpha_2} \cos{\beta_2} \cos({\beta_1-\alpha_1 }) \,+\, \sin{\alpha_2} \sin{\beta_2},
  \\
     \label{eq:gh1VV-3hdm}
g_{h_1VV}\ =&\cos\alpha \, \cos\beta_2 \sin( \beta_1-\alpha_1)\, + \, \sin\alpha \, \Big[ \cos\alpha_2 \sin\beta_2 
 \, -\, \cos\beta_2 \sin\alpha_2\cos(\beta_1-\alpha_1)\,\Big], 
  \\
  \label{eq:gh2VV-3hdm}
  g_{h_2VV}\ =&\ \cos\alpha\, \Big[ \cos\alpha_2 \sin\beta_2-\cos\beta_2\sin\alpha_2\cos(\beta_1-\alpha_1)\,\Big]\,-\,\sin\alpha \, \cos\beta_2 \sin(\beta_1-\alpha_1).
\end{align}
Evidently, there are three possible scenarios for which the 125-GeV resonance
can be identified with the SM-like Higgs boson \cite{Darvishi:2021txa}. Here, we consider the 
{\em canonical} SM-like Higgs scenario i.e. $\beta_1= \alpha_1$ and $\beta_2=\alpha_2$, where $M_H \approx 125$~GeV with the coupling strength $g_{HVV} = 1$ and $g_{h_{1,2}VV} =0$.

In the Higgs basis $\{ H_{1,2,3}\}$, the $3 \times 3$ mass matrix of the CP-even scalars
is given by
 \begin{align}
\widehat{\mathcal{M}}_{\rm S}^2\,=\,\left(
\begin{array}{ccc}
    \label{mass-wtoH}
\widehat{A} & \widehat{C}_{1} & \widehat{C}_{2}
\\
\widehat{C}_{1} & \widehat{B}_{1}& \widehat{C}_{3}
\\
\widehat{C}_{2} & \widehat{C}_{3} &\widehat{B}_{2}
\\
\end{array}
\right)\,=\,\mathcal{O}_\beta\,
\mathcal{M}_{\rm S}^{2} \,  \mathcal{O}^{\sf T}_\beta\,.
\end{align}
The details of the above matrix are presented in \cite{Darvishi:2021txa}.

In the SM alignment limit $\beta_1=\alpha_1$ and $\beta_2=\alpha_2$ under consideration,
the mass parameter $\widehat{A}$ becomes equal to~$M_H^2$, while the parameters $\widehat{C}_1$ and $\widehat{C}_2$ vanish. Thus, taking the limit $\widehat{C}_{1,2} \to 0$, the following relationships between the quartic couplings may be derived:
 \begin{align}
  \label{C1,C2=0}
&\lambda_{11}=\lambda_{22}=\lambda_{33}=(\lambda_{1122}+\lambda_{1221}+\lambda_{1212})/2,\quad \lambda_{1122}=\lambda_{1133}=\lambda_{2233}, \nonumber \\ &\lambda_{1221}=\lambda_{1331}=\lambda_{2332},\quad \lambda_{1212}=\lambda_{1313}=\lambda_{2323},
\end{align}
while the remaining quartic couplings are zero. Note that by virtue of symmetries identified in \eqref{NAS} with $n=3$ the above relationships between the quartic couplings can be met.

The reduced $H$-boson couplings to the EW gauge bosons in a power expansion of $\theta_{\mathcal{S}_{1,2}}\equiv {\widehat{C}_{1,2}/\widehat{B}_{1,2}}$ and $\theta_{\mathcal{S}_{12}}\equiv {\widehat{C}_{1}/\widehat{B}_{2}}$, are given by the following approximate analytic expressions:
\begin{align}
	g_{HVV}\ &\simeq\ 1\, -\,  \theta_{\mathcal{S}_{1}}^2/4 ,
\nonumber \\
	g_{h_1VV}\ &\simeq\ c_{\alpha} c_{\beta_2}\, \theta_{\mathcal{S}_{1}} 
	+ s_{\alpha} \, \theta_{\mathcal{S}_{2}} \,
	(1+\theta_{\mathcal{S}_{1}})/4,
\nonumber \\                                                                                                                                                	g_{h_2VV}\ &\simeq\  - s_{\alpha}c_{\beta_2}\, \theta_{\mathcal{S}_{1}} 
	+ c_{\alpha}   \theta_{\mathcal{S}_{2}} \,(1+\theta_{\mathcal{S}_{1}})/4 .
\end{align}
Likewise, to order $\widehat{C}_{1,2}^2/\widehat{B}^2_{1,2}$, the SM-normalised couplings of the $H$ boson
to fermions are dictated by the following approximate analytic formulae:
\begin{align}
	g_{H d \bar{d}}\ &\simeq\ 1- (t_{\beta_1} + t_{\beta_2}) \; \theta_{\mathcal{S}_{1}} 
 	-(1-t_{\beta_1}  t_{\beta_2}) \; \theta_{\mathcal{S}_{2}}^2 ,
\nonumber \\
	g_{H u \bar{u}}\ &\simeq\, 1 - t_{\beta_1}^{-1}\; \theta_{\mathcal{S}_{1}}
	+ {t_{\beta_2} \over 2} (1 - t_{\beta_1}^{-1} \; \theta_{\mathcal{S}_{1}}) \; \theta_{\mathcal{S}_{2}}\,-\, { 1 \over 8}( 4 \theta_{\mathcal{S}_{1}}^2 + \theta_{\mathcal{S}_{2}}^2),
\nonumber \\
 g_{H l \bar{l}}\ &\simeq\, 1 - 2 \theta_{\mathcal{S}_{2}} t_{\beta_2}^{-1} (4 +  \theta_{\mathcal{S}_{1}} \; \theta_{\mathcal{S}_{12}})^{-1}
-{2 \; \theta_{\mathcal{S}_{2}}^2 }(4 +  \theta_{\mathcal{S}_{1}}\; \theta_{\mathcal{S}_{12}} )^{-2}.
\end{align}
We note that the analytic expressions of the reduced $H$-boson couplings to the gauge bosons and fermions go to the SM value $1$, when the exact SM alignment limit~$\widehat{C}_{1,2}\,\to 0$ is considered, or when the new-physics mass scales $\widehat{B}_{1,2} \sim M^2_{h_{1,2}}$ are taken to infinity.  Since we are interested in the former possibility which in turn implies a richer collider phenomenology, we will study scenarios in which SM alignment limit is accomplished by virtue of~$\mathrm{Sp}(6)$~symmetry.

\section{Maximally Symmetric {\boldmath$n$}HDMs}

MS-$n$HDMs are the most economic settings that realise naturally SM
alignment from the general class of $n$HDMs.  The
$\mathrm{Sp}(2n)$-invariant potential has the same form as the SM
potential, \bea V_{\rm SM}=-m^2 \left|\Phi\right|^2\,+\, \lambda
\left|\Phi\right|^4, \eea that contains a single mass term and a
single quartic coupling.  In the case of the 2HDM
$\mathrm{Sp}(4){/Z_2}$-invariant potential, the so called MS-2HDM is
given by
\bea V_{\rm{MS-2HDM}}\,=\, - m^2 \left( \left| \Phi_1
  \right|^2\, +\, \left| \Phi_2 \right|^2 \right) + \lambda \left(
  \left| \Phi_1 \right|^2 \,+\, \left| \Phi_2 \right|^2 \right)^2,
  \label{eq:VMS2HDM}
\eea
where the non-zero parameters have the following relations,
\begin{align}
&m^2\,=\,m_{11}^2\,=\,m_{22}^2, \qquad \lambda\, =\, \lambda_{1}\, =\, \lambda_{2} =\, \lambda_{3}/2.
\nonumber
\label{eq:m-c}
\end{align}
Likewise, in the case of $\mathrm{Sp}(6){/Z_2}$-invariant potential, the so called MS-3HDM we have
\begin{equation}
V_{\rm MS-3HDM}\ =\ -\, m^2\, \Big( \left| \Phi_1 \right|^2+
  \left| \Phi_2 \right|^2 + \left| \Phi_3 \right|^2 \Big)\: 
 +\: \lambda\, \Big( \left| \Phi_1 \right|^2 + \left| \Phi_2 \right|^2+ \left| \Phi_3 \right|^2
   \Big)^2, 
   \label{ms-3hdm-v0}
\end{equation}
with the following relationship between non-zero parameters:
\begin{align}
m^2\,=\,m_{11}^2\,=\,m_{22}^2\,=\,m_{33}^2\,,\qquad \, \lambda\, =\, \lambda_{11}\, =\, \lambda_{22}\, =\, \lambda_{33}\, =\, \lambda_{1122}/2\, =\, \lambda_{1133}/2\, =\, \lambda_{2233}/2\,,
\nonumber
\end{align}
In particular, in the MS-3HDM one can realize the mixing angles that diagonalise
the heavy sectors of the CP-even, CP-odd and charged- scalar mass matrices are all equal, i.e. $
\bm{\alpha\, =\, \rho\, =\, \sigma}$,
that is a distinct and unique feature of the MS-3HDM.

Moreover, after SSB and in the Born approximation, the MS-$n$HDM predicts one CP-even scalar $H$ with non-zero squared mass $M^2_H\,= 2 \lambda v^2$, while all other\,scalars are massless pseudo-Goldstone bosons with sizeable gauge and Yukawa interactions. Therefore, we need to consider two main sources that break the $\mathrm{Sp}(2n)$ symmetry of the theory: (i)~the RG effects of the gauge and~Yukawa couplings on the potential parameters, and (ii)~soft symmetry-breaking bilinear masses, rendering all pseudo-Goldstone fields sufficiently heavy in agreement with current LHC data and other low-energy experiments. Hence, the complete MS-$n$HDM potential is
\begin{align}
   \label{eq:V+DV}
   V\ &=\ V_{{\rm MS}-n{\rm HDM}}\: +\: \displaystyle\sum_{i,j=1}^n\, m_{ij}^2\, ( \Phi_i^{\dagger} \Phi_j)\,.
\end{align}
Henceforth, with this minimal addition, the
scalar-boson masses in MS-2HDM are given by
\begin{align}
  \label{eq:m12}
 M_H^{2} = 2\lambda_2 v^{2}, \qquad M_h^{2}\ =\ M_a^{2}\ =\ M_{h^{\pm}}^{2}\ 
 =\ {\text{Re}(m_{12}^{2}) \over c_{\beta} s_{\beta} },
\end{align}
where this degeneracy will be broken by RG effect. 

In the case of MS-3HDM, the scalar masses are found to be:
\begin{align}
    \label{eq:mij}
 M_H^2 \, \simeq\ 2\lambda_{22} v^2, \quad
 M_{h_1,h_2}^2\, \simeq \, M_{a_1,a_2}^2\ \simeq\ M_{{h_1^{\pm},h_2^{\pm}}}^2\ \simeq\  {1 \over 2}\bigg[\,S_1 + S_2\mp\big[{(S_1- S_2 )^2+4 S_3^2}\big]^{1/2}\,\bigg]\;,
\end{align}
with 
\begin{align}
  S_1= \frac{{m_{12}^2}+ { t_{\beta_2}}\big({{m_{13}^2 \, s_{\beta_1}^3} + m_{23}^2 \, c_{\beta_1}^3 }\big)}{c_{\beta_1} s_{\beta_1}}, \quad
S_2= \frac{m_{13}^2 \, c_{\beta_1}+m_{23}^2 \, s_{\beta_1} }{c_{\beta_2} s_{\beta_2}}\, \;,
\quad
S_3= \frac{m_{13}^2 \, s_{\beta_1}-m_{23}^2 \, c_{\beta_1}}{c_{\beta_2}}\;.
\end{align}
Therefore, with the introduction of soft symmetry-breaking bilinears, all pseudo-Goldstone bosons, $h_{1,2}$, $a_{1,2}$ and $h_{1,2}^\pm$, receive appreciable masses at the tree level. 

\subsection{Quartic Coupling Unification}\label{QCU}
The interesting feature of MS-$n$HDM is quartic coupling unification, where all quartic couplings can unify to a single value $\lambda$ at very high-energy scales~$\mu_X$. As the highest scale of unification, we take the values at which $\lambda (\mu_X) = 0$. We will see that there are two such conformally invariant unification points in the MS-$n$HDM which we distinguish them as~$\mu^{(1,2)}_X$, with $\mu^{(1)}_X \le \mu^{(2)}_X$. 

\begin{figure}
\centering
\includegraphics[width=0.495\textwidth]{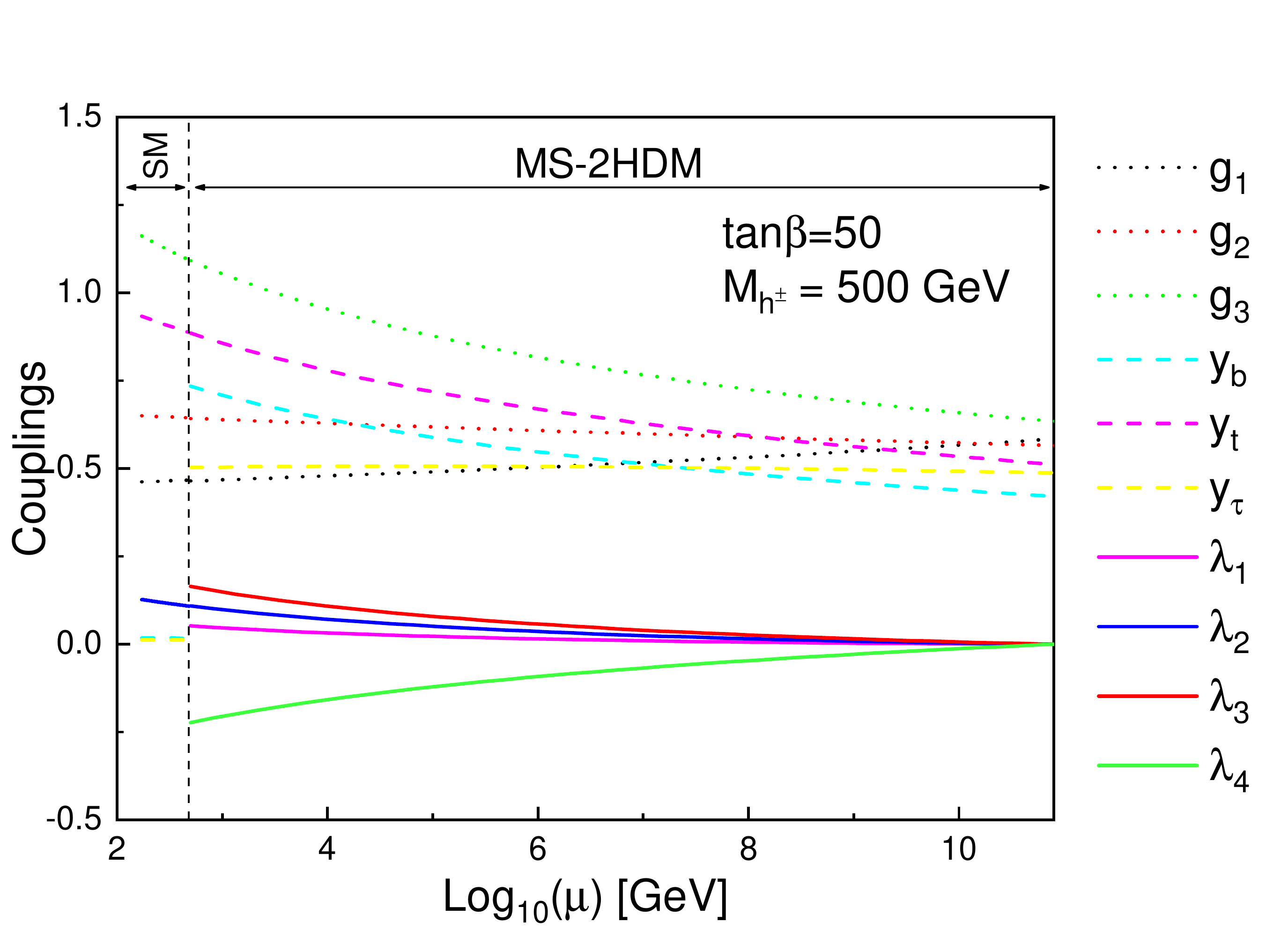}
\includegraphics[width=0.495\textwidth]{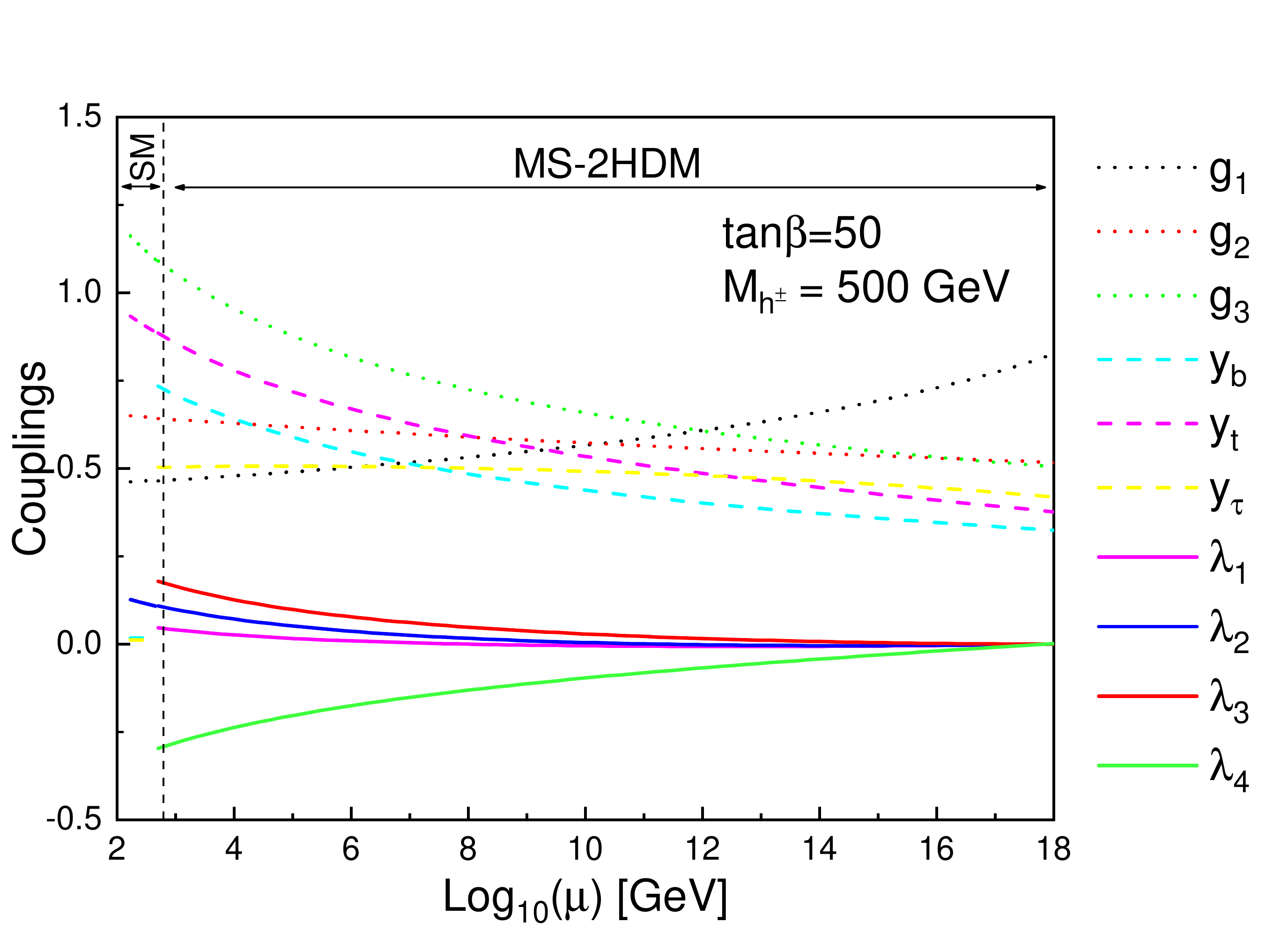}
\caption{\it In the left (right), the RG evolution of the quartic couplings in MS-3HDM from the threshold 
scale $M_{h^\pm} = 500\,\mathrm{GeV}$ up to their first (second) quartic coupling unification scale~${\mu^{(1)}_X \sim 10^{11}}\,\mathrm{GeV}$ ($\mu^{(2)}_X \sim 10^{18}\,\mathrm{GeV}$) for $\tan\beta= 50$.}
\label{first-MS2HDM}
\end{figure}

\begin{figure}
\centering
\includegraphics[width=0.49\textwidth]{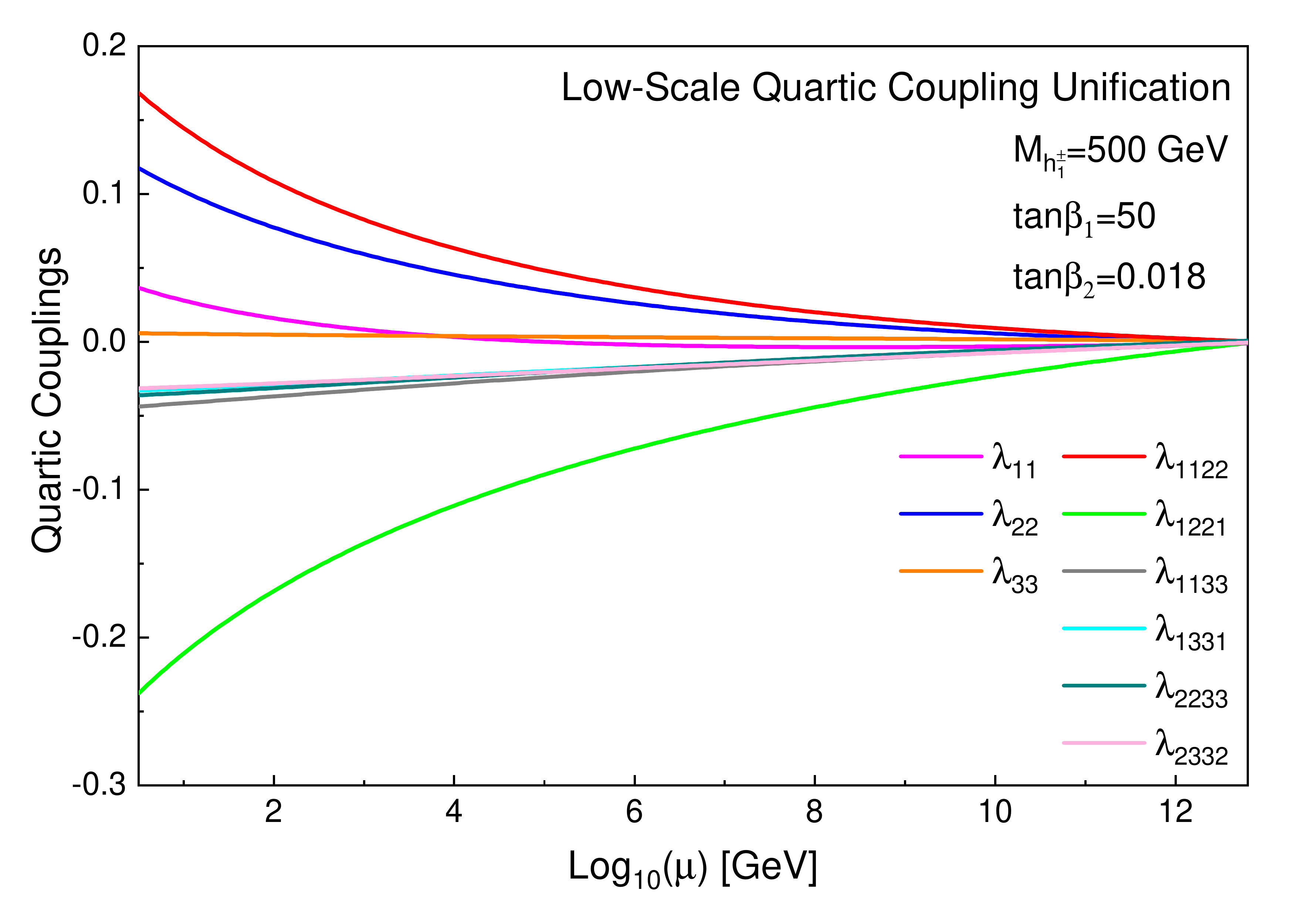}
\includegraphics[width=0.49\textwidth]{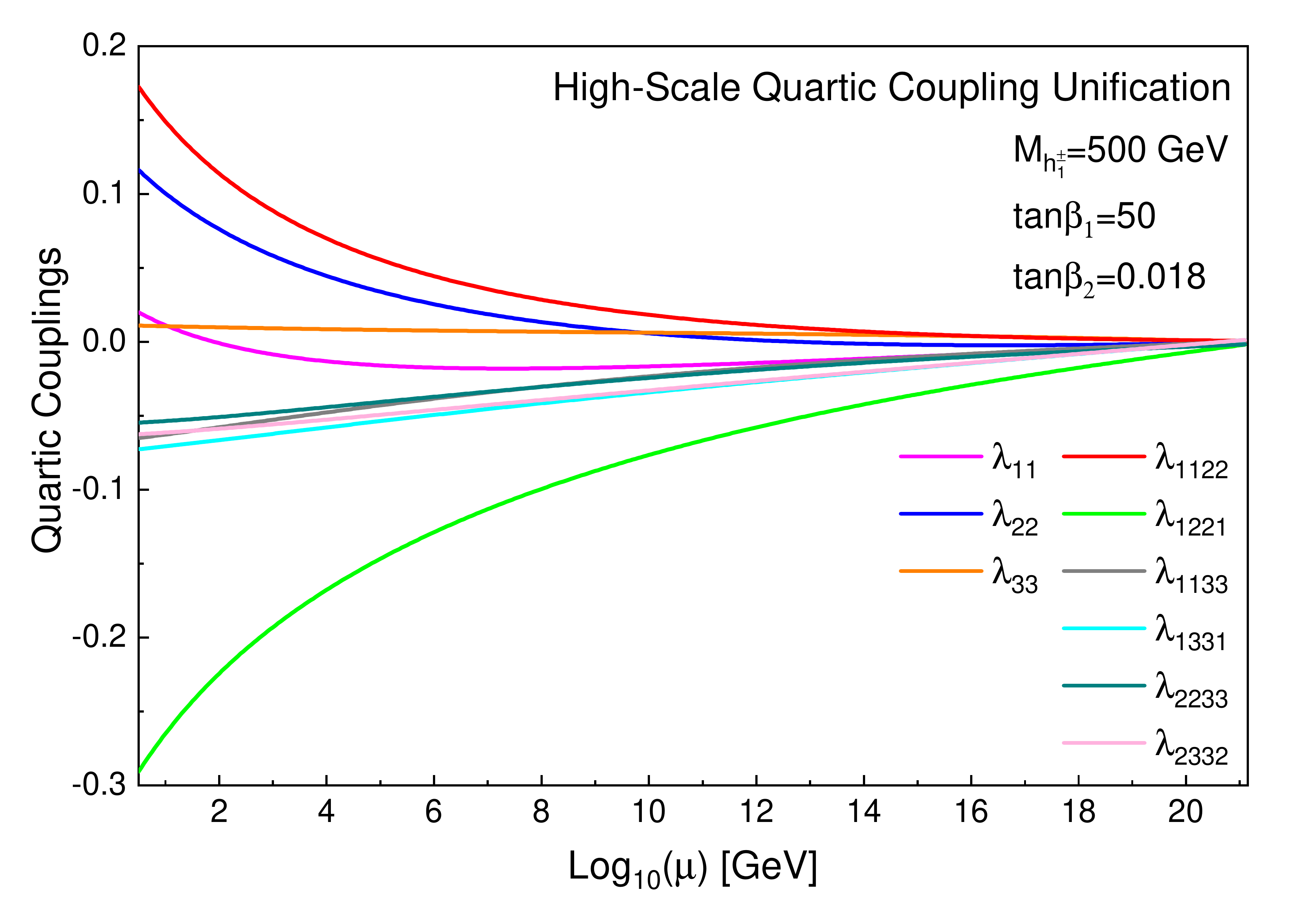}
\caption{\it In the left (right), the RG evolution of the quartic couplings in MS-3HDM from the threshold 
scale $M_{h_1^\pm} = 500\,\mathrm{GeV}$ up to their first quartic coupling unification scale~${\mu^{(1)}_X \sim 10^{13}}\,\mathrm{GeV}$ ($\mu^{(2)}_X \sim 10^{21}\,\mathrm{GeV}$) for $\tan\beta_1= 50$ and $\tan\beta_2 =0.018$.}
\label{first-MS3HDM}
\end{figure}

In Figures~\ref{first-MS2HDM} and ~\ref{first-MS3HDM}, we display the RG evolution of all quartic couplings for MS-2HDM and MS-3HDM, with a low charged Higgs mass $M_{h^\pm}=M_{h_1^\pm} = 500$~GeV. From the left panels of these figures, we observe that in the case of MS-3HDM (MS-2HDM) the quartic coupling $\lambda_{22}$ ($\lambda_{2}$), which determines the SM-like Higgs-boson mass~$M_H$, decreases at high RG scales, due to the running of the top-Yukawa coupling~$y_t$. The coupling $\lambda_{22}$ ($\lambda_{2}$) turns negative just above the quartic coupling unification scale $\mu^{(1)}_X \sim 10^{13}$\,GeV ($\mu^{(1)}_X \sim 10^{11}$\,GeV), at which all quartic couplings vanish. Below $\mu^{(1)}_X$, the MS-3HDM (MS-2HDM) quartic couplings exhibit different RG runnings, and especially the couplings $\lambda_{ijij}$ (with $i\neq j$) take on non-zero values. From the right panels of Figures~\ref{first-MS2HDM} and ~\ref{first-MS3HDM}, we observe that in addition to the conformal unification point $\mu^{(1)}_X$, there is in general a second and higher conformal point $\mu^{(2)}_X \sim 10^{21}\,\mathrm{GeV}$ ($\mu^{(2)}_X \sim 10^{18}\,\mathrm{GeV}$) in the MS-3HDM (MS-2HDM).  
  
\begin{figure}[t]
\centering
\includegraphics[width=0.49\textwidth]{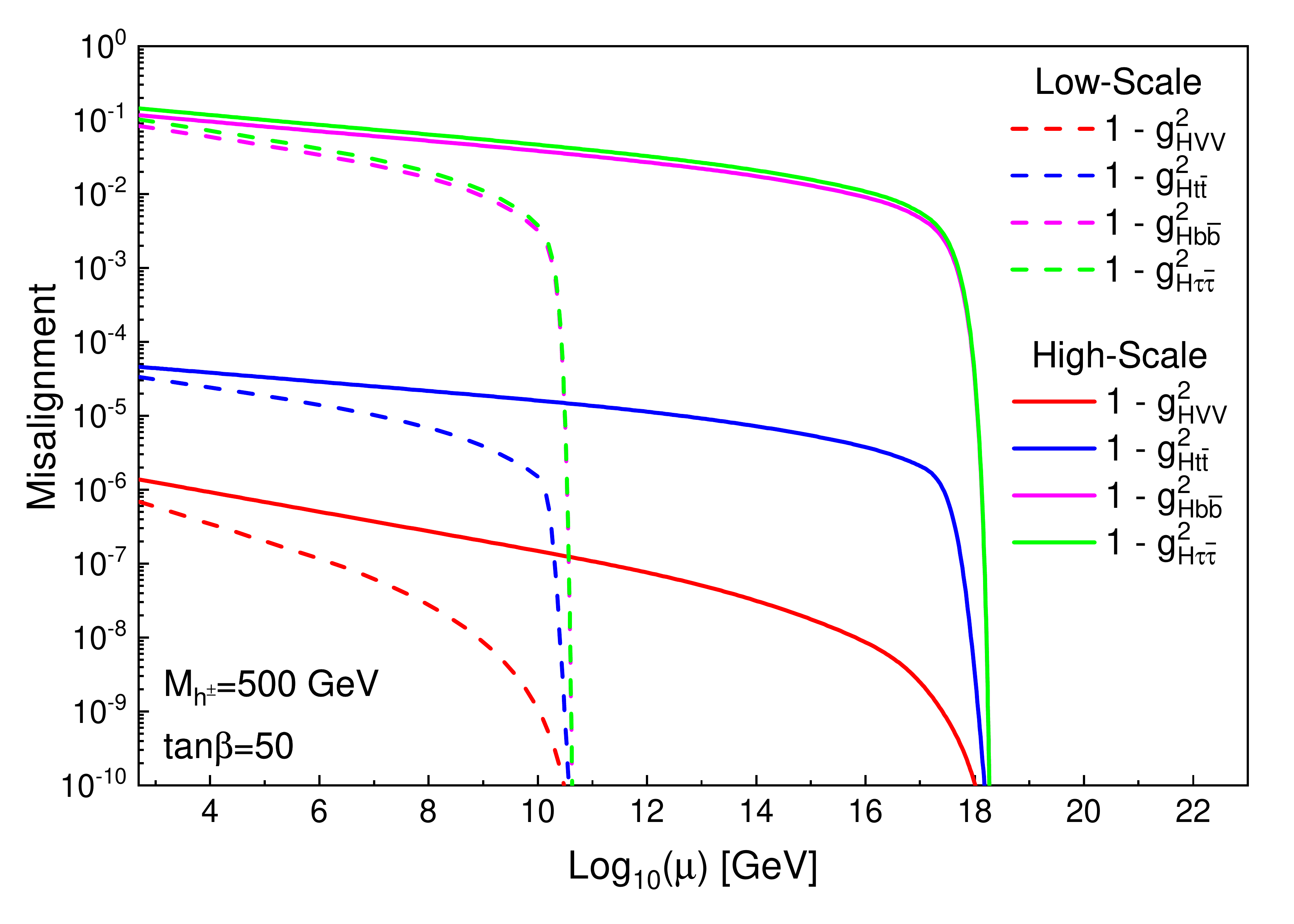}
\includegraphics[width=0.49\textwidth]{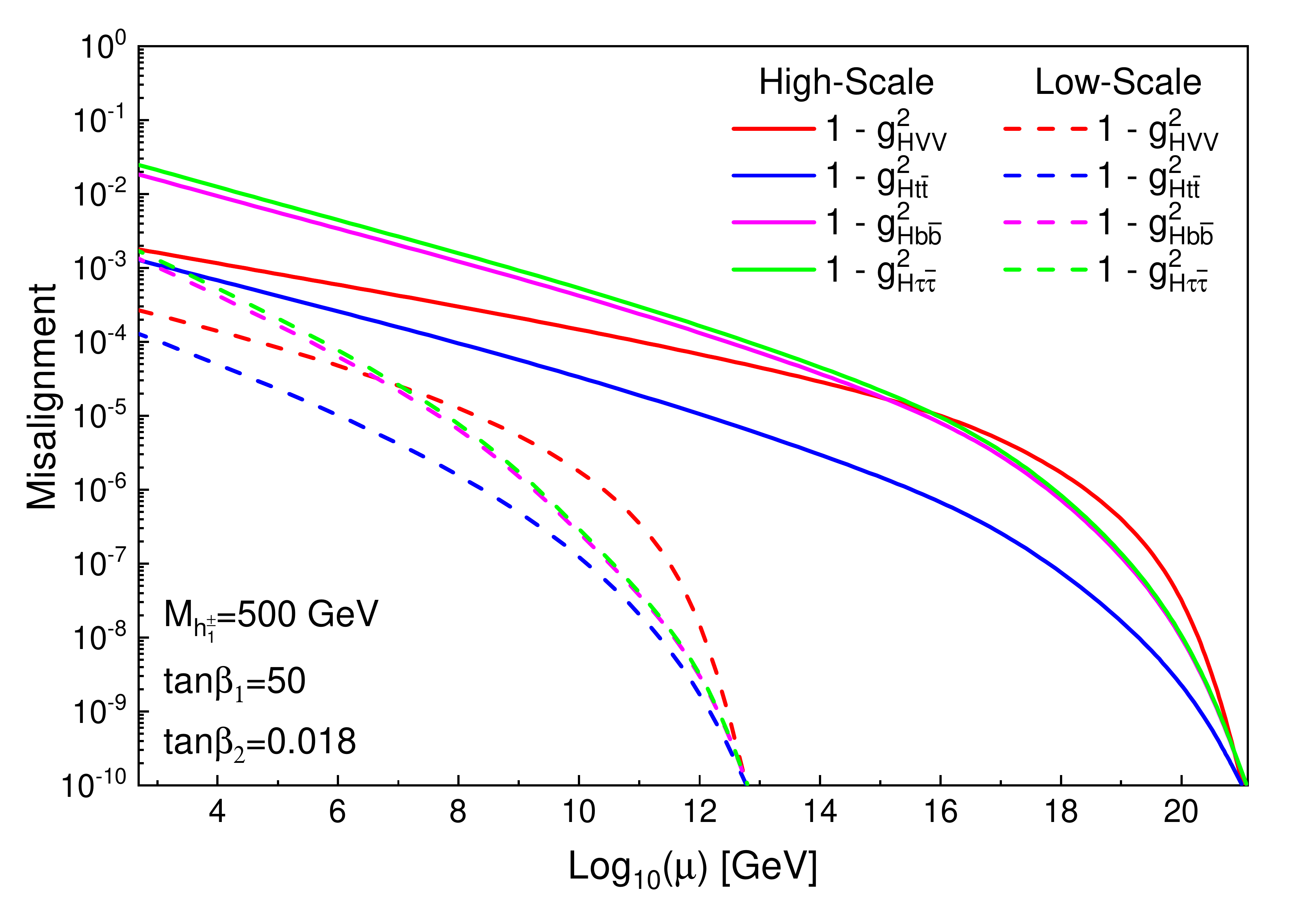}
\caption{\it Numerical estimates of the misalignment parameter $1 - g^2_{HXX}$ pertinent to the $HXX$-coupling (with $X=V,t,b,\tau$ and $V= W^\pm,Z$) as functions of the RG scale~$\mu$, for the low-scale and the high-scale quartic coupling unification scenarios in MS-2HDM (left panel) and MS-3HDM (right panel).}
\label{MisAlign-MS3HDM}
\end{figure}

  We have already seen how at the conformal points, $\mu^{(1)}_X$ and $\mu^{(2)}_X$, all quartic couplings vanish simultaneously, leading to an exact SM alignment. Nevertheless, for lower RG scales, the maximal symmetry is broken, resulting in calculable non-vanishing misalignment predictions. In Figure~\ref{MisAlign-MS3HDM}, we present our numerical estimates of the predicted deviations for the SM-like Higgs-boson coupling $HXX$ (with $X=W^\pm,Z,t,b,\tau $) from its respective SM value in MS-2HDM (left panel) and in MS-3HDM (right panel). Specifically, Figure~\ref{MisAlign-MS3HDM} exhibits the dependence of the misalignment parameter $1-g^2_{HXX}$ (with $g_{H_{\mathrm{SM}}XX}\,=\,1$) as functions of the RG scale~$\mu$, for both low-\,and\,high-scale quartic coupling unification scenarios.  We observe that the normalised couplings $g_{HVV}$ and $g_{Htt}$ reach their SM values $g_{H_{\mathrm{SM}}V V} \,=\, g_{H_{\mathrm{SM}}tt}  \,=\, 1$ at the two quartic coupling unification points, $\mu^{(1)}_X$ and $\mu^{(2)}_X$. Moreover, the deviation of ~$g_{Hbb}$ and ~$g_{H\tau \tau}$ from their SM values get larger for the higher-scale unification scenario and can be fitted to the observed data within the 68$\%$ CL~\cite{201606,Palmer:2021gmo}.

\section{Conclusions} \label{con}

We have analysed the basic low-energy structure of the general 2HDM and 3HDM. Our study was focused on the canonical SM-like Higgs scenario of the Type-II 2HDM and the Type-V 3HDM, for which conditions on these models parameters for achieving exact SM alignment were derived. Interestingly enough, there are three continuous symmetries which, when imposed on the $n$HDM scalar potential, are sufficient to ensure SM alignment: (i)~$\mathrm{Sp}(2n)$, (ii)~$\mathrm{SU}(n)_{\rm HF}$,
and (iii)~$\mathrm{SO(n})_{\rm HF}$. Amongst these symmetries, the most economic setting is the Maximally Symmetric multi-Higgs Doublet Model (MS-$n$HDM), whose potential obeys an $\mathrm{Sp(2}n)$ symmetry.  This symmetry is softly broken by bilinear masses $m^2_{ij}$ (with $i\neq j$), as well as explicitly by hyper\-charge and Yukawa couplings through RG effects, whilst the theory~exhibits quartic coupling unification up~to~the Planck scale. In particular, we have shown that all quartic couplings in the MS-3HDM (MS-2HDM) can unify at high-energy scales~$\mu_X$ and vanish simultaneously at two distinct conformal points, that are denoted by~$\mu^{(1,2)}_X$ with $\mu_X^{(1)} \lesssim 10^{13}\,\text{GeV}$ ($\mu_X^{(1)} \lesssim 10^{11}\,\text{GeV}$) and $\mu_X^{(2)} \gtrsim 10^{21}\,\text{GeV}$ ($\mu_X^{(2)} \gtrsim 10^{18}\,\text{GeV}$).

The MS-$n$HDM is a remarkably predictive scenario, as it only depends
on a few theoretical parameters when compared to the large number of
independent parameters that are required in the general $n$HDM. For
example, in the MS-2HDM the only additional parameters are: (i) the
charged Higgs mass $M_{h^{\pm}}$ and (ii) $\tan\beta$. In the case of
MS-3HDM, besides the ratios of the Higgs-doublet VEVs,
$\tan\beta_{1,2}$, the model is mainly governed by only three input
parameters: the masses of the two charged Higgs bosons,
$M_{h_{1,2}^{\pm}}$, and their mixing angle~$\sigma$. Most notably,
with the help of these input parameters, we have obtained misalignment
predictions for the entire scalar mass spectrum of the theory,
including the interactions of all Higgs particles to the SM fields.
The present results for the
MS-$n$HDM~\cite{Darvishi:2019ltl,Darvishi:2021txa}, demonstrate the
high predictive power of maximally symmetric settings in $n$HDMs. Such
settings not only can naturally provide the experimentally favoured SM
alignment, but also allow us to obtain sharp predictions for the
entire scalar mass spectrum of the theory. In~conclusion${}$, the MS-$n$HDM
provides a unique framework that can be tested or even falsified at the
Large Hadron Collider (LHC) and at upcoming high-energy colliders, such as
the projected Future Circular Collider (FCC) at CERN.

 \section*{Acknowledgements}
\noindent
The work of AP is supported in part by the
Lancaster–Manchester–Sheffield Consortium\- for Fundamental Physics,
under STFC research grant ST/P000800/1. 
\\
The work of MRM is supported under STFC research grant ST/P001246/1. 
\\
The work of ND is supported by the National Natural Science Foundation of China (NSFC)
under grants No. 12022514, No. 11875003 and No. 12047503, and CAS Project for Young
Scientists in Basic Research YSBR-006, by the Development Program of China under Grant
No. 2020YFC2201501 (2021/12/28) and by the CAS President’s International Fellowship
Initiative (PIFI) grant. ND is also supported by the Polish National Science Centre HARMONIA grant under contract UMO-2015/20/M/ST2/00518 (2016-2021).

\end{document}